\documentclass{pinchcr}

\usepackage{amsfonts}

\renewcommand{\Re}{{\mathrm{Re}}\, }

\newcommand{\beq}{\begin{equation}}
\newcommand{\eeq}{\end{equation}}

\newcommand{\ket}[1]{|#1\big>}
\newcommand{\barr}{\begin{eqnarray}}
\newcommand{\earr}{\end{eqnarray}}

\newcommand{\ZZ}{\mathbb{Z}}
\newcommand{\CC}{\mathbb{C}}

\newcommand{\mf}[1]{\mathfrak{#1}}

\begin{document}

\title{Topological phases and quantum computation}

\author{A. Kitaev}

\affiliation{\emph{California Institute of Technology}, Pasadena, CA 91125}

\author{C. Laumann}

\affiliation{\emph{Department of Physics}, \emph{Princeton University}, Princeton NJ 08544}

\authors{2}

\maketitle




\tableofcontents

\maintext

\chapter{Introduction: The quest for protected qubits}
\label{sec:intr-motiv}

\begin{figure}
 \centering
 \includegraphics{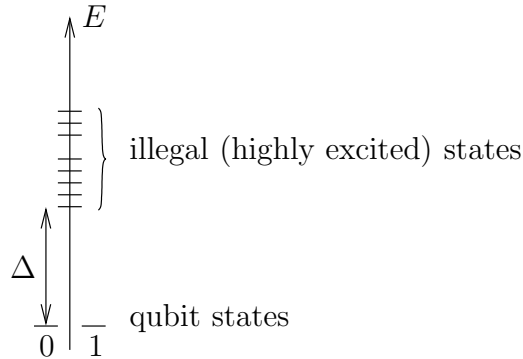}
 \caption{Spectrum of a physical qubit system.}
 \label{fig:qubit-spec}
\end{figure}

The basic building block of quantum computation is the qubit, a system
with two (nearly) degenerate states that can be used to encode quantum
information. Real systems typically have a full spectrum of
excitations that are considered illegal from the point of view of a
computation, and lead to decoherence if they couple too strongly into
the qubit states during some process. See
Fig.~\ref{fig:qubit-spec}. The essential problem then is to preserve
the quantum state of the qubit as long as possible to allow time for
computations to take place.

Assuming the gap $\Delta$ to the illegal states is reasonable, we can quite
generally describe the dynamics of the qubit state by an effective
Sch\"odinger equation
\begin{eqnarray}
 \label{eq:76}
 \frac{d}{dt}\ket{\Psi} = -i H_{\mathrm{eff}}\ket{\Psi}
\end{eqnarray}
where $H_{\mathrm{eff}}$ is the effective qubit Hamiltonian. In
quantum optics, $H_{\mathrm{eff}}$ is often known with high
precision. This is not so in condensed matter systems such as quantum
dots. Even worse, $H_{\mathrm{eff}}$ may fluctuate or include
interaction with the environment. This causes decoherence of the qubit
state.

Ideally, we would like to arrange for $H_{\mathrm{eff}}$ to be zero
(or $H_{\mathrm{eff}} = \epsilon I$) for some good reason. Usually, we
use a symmetry to protect degeneracies in quantum systems. For
example, a quantum spin $\frac{1}{2}$ has a two-fold degeneracy
protected by the $SU(2)$ symmetry, as do the $2s+1$ degeneracies of
higher spins $s$. Indeed, any \emph{non-Abelian} symmetry would
work. Unfortunately, the $SU(2)$ symmetry of a spin is lifted by
magnetic fields and it's generally difficult to get rid of stray
fields.

Rather than symmetry, in what follows we will look to topology to
provide us with physically protected degeneracies in quantum
systems. In particular, we will examine a number of exactly solvable
models in one and two dimensions which exhibit topologically phases --
that is, gapped phases with a protected ground state degeneracy
dependent on the topology of the manifold in which the quantum model
is embedded. In Sec.~\ref{sec:topol-phen-1d} we warm up with the
study of several quantum chains that exhibit Majorana edge modes and
thus a two-fold degeneracy on open chains. The topological phenomena
available in two dimensional models are much richer and will be the
focus of the remaining three sections. We introduce and solve the
toric code on the square lattice in
Sec.~\ref{sec:two-dimens-mathbbz_2}, exhibiting its topolological
degeneracy and excitation spectrum explicitly. The following section
steps back to examine the general phenomenology of quasiparticle
statistics braiding in two dimensional models. Finally, in
Sec.~\ref{sec:honeyc-latt-model} we introduce the honeycomb lattice
model which exhibits several kinds of topological phases, including
that of the simple toric code and, in the presence of time reversal
symmetry breaking, a gapped phase with chiral edge modes protected by
the topology of the Fermi surface.

\chapter{Topological phenomena in 1D: boundary modes in the Majorana chain}
\label{sec:topol-phen-1d}

We will consider two examples of 1D models with $\ZZ_2$ symmetry and
topological degeneracy: the \emph{transverse field Ising model} (TFIM)
and the \emph{spin-polarized superconductor} (SPSC).  Although these
models look rather different physically, we will find that they are
mathematically equivalent and that they both exhibit a topological
phase in which the ground state degeneracy is dependent on the
boundary conditions of the chain. That is, the ground state on an open
chain is 2-fold degenerate due to the presence of boundary zero modes,
whereas the ground state is unique on a closed loop. This topological
degeneracy will be stable to small \emph{local} perturbations that
respect the $\ZZ_2$ symmetry. More details on these models may be
found in \shortciteN{Kitaev:2000p5909}.

\begin{enumerate}
\item
 \label{ex:tfim}
 The \emph{transverse field Ising model} is a spin-1/2 model with
 Hamiltonian:
 \begin{equation}
   \label{eq:tfim-ham}
   H_S = -J \sum_{j=1}^{N-1} \sigma_j^x \sigma_{j+1}^x - h_z
   \sum_{j=1}^N \sigma_j^z.
 \end{equation}
 Here $J$ is the ferromagnetic exchange coupling in the $x$ direction
 and $h_z$ is a uniform transverse ($z$) field. This model has a
 $\ZZ_2$ symmetry given by a global spin flip in the $\sigma_x$ basis:
 \begin{equation}
   \label{eq:tfim-sym}
   P_S = \prod_{j=1}^N \sigma_j^z
 \end{equation}

\item
 \label{ex:1dsc}
 The \emph{spin-polarized 1-D superconductor} is a fermionic system
 with Hamiltonian:
 \begin{eqnarray}
   \label{eq:1dsc}
   H_F &=& \sum_{j=1}^{N-1}\left(-w (a^\dagger_ja_{j+1} +
     a^\dagger_{j+1}a_j) 
     + \Delta a_j a_{j+1} + \Delta^*
     a^\dagger_{j+1}a^\dagger_j\right) \nonumber\\
   &&- \mu \sum_{j=1}^N\left(a^\dagger_j a_j - \frac{1}{2}\right)
 \end{eqnarray}
 where $a_j$ and $a^\dagger_j$ are fermionic annihilation and
 creation operators, $w$ is the hopping amplitude, $\Delta$ is the
 superconducting gap and $\mu$ is the chemical potential. For
 simplicity, we will assume that $\Delta = \Delta^* = w$, so that
 \begin{equation}
   \label{eq:1dsc-simp}
   H_F = w \sum_{j=1}^{N-1}(a_j - a_j^\dagger)(a_{j+1} +
   a_{j+1}^\dagger) 
   - \mu \sum_{j=1}^N\left(a_j^\dagger a_j - 1/2\right).
 \end{equation}
 This model has a $\ZZ_2$ symmetry given by the fermionic parity operator:
 \begin{equation}
   \label{eq:1dsc-sym}
   P_F = (-1)^{\sum_j a_j^\dagger a_j}
 \end{equation}
\end{enumerate}

Although the two models are mathematically equivalent, as we will see
in Sec. \ref{sec:reduction-tfim-1dsc}, they are clearly physically
different. In particular, for the superconductor, the $\ZZ_2$ symmetry
of fermionic parity cannot be lifted by any local physical operator,
as such operators must contain an even number of fermion
operators. Unfortunately, for the spin system the degeneracy is lifted
by a simple longitudinal magnetic field $h_x \sum_j \sigma_j^x$ and
thus the topological phase of the TFIM would be much harder to find in
nature.

\section{Nature of topological degeneracy (spin language)}
\label{sec:nature-topol-degen}

\newcommand{\larr}{\leftarrow}
\newcommand{\rarr}{\rightarrow}

Consider the transverse field Ising model of
Eq. (\ref{eq:tfim-ham}). With no applied field, there are a pair
of Ising ground states ($h_z=0$):
\begin{eqnarray}
 \label{eq:tfim-gs}
 \ket{\psi_\rarr} & = &\ket{\rarr\rarr\rarr\cdots\rarr},~
 \ket{\psi_\larr} = \ket{\larr\larr\larr\cdots\larr}.
\end{eqnarray}
The introduction of a small field $h_z$ allows the spins to flip in the
$\sigma^x$ basis. In particular, tunneling between the two classical
ground states arises via a soliton (domain-wall) propagating from one
side of the system to the other: 
\begin{eqnarray}
 \label{eq:tfim-soliton}
 \ket{\rarr\rarr\rarr\cdots\rarr} &\longrightarrow&
 \ket{\larr:\rarr\rarr\cdots\rarr} \longrightarrow
 \ket{\larr\larr:\rarr\cdots\rarr} \\
 &\longrightarrow& 
 \ket{\larr\larr\larr:\cdots\rarr} 
 \longrightarrow \cdots
 \longrightarrow
 \ket{\larr\larr\larr\cdots\larr}.
\end{eqnarray}
As usual, the tunneling amplitude $t$ associated with this transition 
falls off exponentially in the distance the soliton must propagate
\begin{equation}
 \label{eq:tfim-tunnel}
 t \sim e^{-N/\xi}
\end{equation}
where $\xi$ is the correlation length of the model. The two-fold
degeneracy is therefore lifted by the effective Hamiltonian:
\begin{equation}
 \label{eq:tfim-eff}
 H_{\mathrm{eff}} = \left( \begin{array}{cc}
0 & -t \\
-t & 0 \end{array}\right).
\end{equation}
The splitting is exponentially small in the system size and
the two-fold degeneracy is recovered in the thermodynamic
limit as expected. Moreover, it is clear why introduction of a
longitudinal field $h_x$ will fully split the degeneracy.

\section{Reduction of TFIM to SPSC by the Jordan-Wigner transformation}
\label{sec:reduction-tfim-1dsc}

\newcommand{\up}{\uparrow}
\newcommand{\down}{\downarrow}

To show the equivalence of the one dimensional models introduced
above, we will use a standard Jordan-Wigner transformation to convert
the spins of the Ising model into fermions. It is
perhaps not surprising that a fermionic description exists for spin
1/2 systems -- we simplify identify the up and down state of each spin
with the presence or absence of a fermion. The only difficulty arises
in arranging the transformation so that the appropriate
(anti)-commutation relations hold in each description. The
Jordan-Wigner transformation does this by introducing string-like
fermion operators that work out quite nicely in 1-D nearest neighbor models.

To reduce $H_S$ to $H_F$, we
\begin{enumerate}
\item Associate the projection onto the $z$-axis of the spin with the
 fermionic occupation number:
 \begin{equation}
   \label{eq:2}
   \ket{\up} \leftrightarrow n = 0,~~\ket{\down} \leftrightarrow n = 1.
 \end{equation}
 That is,
 \begin{equation}
   \label{eq:3}
   \sigma_j^z = (-1)^{a_j^\dagger a_j}.
 \end{equation}

\item Introduce the string-like annihilation and creation operators
 \begin{eqnarray}
   \label{eq:4}
   a_j &=& \left(\prod_{k=1}^{j-1} \sigma_k^z\right) \sigma_j^+ \nonumber\\
   a_j^\dagger &=& \left(\prod_{k=1}^{j-1} \sigma_k^z\right) \sigma_j^-
 \end{eqnarray}
 where $\sigma^+$ and $\sigma^-$ are the usual spin raising and lower
 operators. At this stage, we can check that the usual fermionic
 anticommutation relations hold for the $a_j, a_j^\dagger$:
 \begin{equation}
   \label{eq:5}
   \left\{a_i, a_j^\dagger\right\} = \delta_{ij}
 \end{equation}

\item Observe that 
 \begin{equation}
   \label{eq:6}
   \sigma_j^x \sigma_{j+1}^x = - (a_j - a_j^\dagger) (a_{j+1} + a_{j+1}^\dagger),
 \end{equation}
 so $H_S$ (Eq.~(\ref{eq:tfim-ham})) reduces to $H_F$
 (Eq.~(\ref{eq:1dsc-simp})) with 
 \begin{equation}
   \label{eq:80}
   w = J,~~~\mu = -2 h_z
 \end{equation}
\end{enumerate}

\section{Majorana operators}
\label{sec:majorana-operators}

Majorana operators provide a convenient alternative representation of
Fermi systems when the number of particles is only conserved modulo 2,
as in a superconductor. Given a set of $N$ Dirac fermions with
annihilation/creation operators $a_j, a_j^\dagger$, we can define a
set of $2N$ real Majorana fermion operators as follows:
\begin{eqnarray}
 \label{eq:Maj-def}
 c_{2j-1} &=& a_j + a_j^\dagger \nonumber\\
 c_{2j} &=& \frac{a_j - a_j^\dagger}{i}.
\end{eqnarray}
These operators are Hermitian and satisfy a fermionic
anticommutation relation:
\begin{eqnarray}
 \label{eq:8}
 c_k^\dagger &=& c_k \nonumber\\
 c_k^2 &=& 1, ~~ c_k c_l = - c_l c_k (k \ne l).
\end{eqnarray}
Or, more compactly,
\begin{eqnarray}
 \label{eq:1}
 \left\{c_k,c_l\right\} = 2\delta_{kl}.
\end{eqnarray}
From any pair of Majorana operators, we can construct an annihilation
and creation operator for a standard Dirac fermion ($a = (c_1 + i c_2)
/ 2$ and h.c.), and thus the unique irreducible representation for the
pair is a 2-dimensional Hilbert space which is either occupied or
unoccupied by the $a$ fermion.

\begin{figure}
 \centering
 \vspace{0.5cm}
 \includegraphics{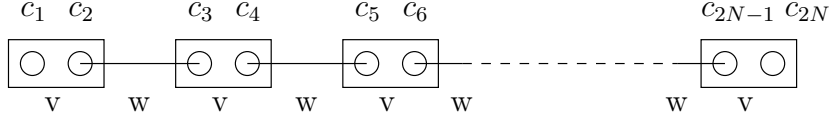}
 \vspace{0.5cm}
 \caption{Majorana chain representation of 1-d superconductor. Each
   boxed pair of Majoranas corresponds to one site of the original
   fermionic chain.}
 \label{fig:major-chain}
\end{figure}

Both models $H_S$ and $H_F$ can be written as 
\begin{equation}
 \label{eq:maj-ham}
 H_{\mathrm{maj}} = \frac{i}{2}\left( v \sum_{j=1}^N c_{2j-1} c_{2j} + w
   \sum_{j=1}^{N-1} c_{2j} c_{2j+1} \right)
\end{equation}
where $v = h_z = - \mu / 2$ and $w = J$. The $\ZZ_2$ symmetry of
fermionic parity is given in the Majorana language by
\begin{equation}
 \label{eq:84}
 P_{\mathrm{maj}} = \prod_{k=1}^N(-i c_{2k-1}c_{2k}).
\end{equation}

We can view this model graphically as a chain of coupled
Majorona modes, two to each of the $N$ sites of the original problem
as in Fig.~\ref{fig:major-chain}.  If $v = 0$, then the Majorana modes
at the ends of the chain are not coupled to anything. This immediately
allows us to identify the 2-fold ground state degeneracy in
$H_{\mathrm{maj}}$ as the tensor factor given by the 2-dimensional
representation of the boundary pair $c_1, c_{2N}$.

We will see in Sec.~\ref{sec:gener-prop-quadr} that if $v \ne
0$ but $|v| < w$, the operators $c_1$ and $c_{2N}$ are replaced by
some \emph{boundary mode operators} $b_l$, $b_r$. The effective
Hamiltonian for this piece of the system is then
\begin{equation}
 \label{eq:9}
 H_{\mathrm{eff}} = \frac{i}{2}\epsilon b_l b_r = \epsilon(a^\dagger a - \frac{1}{2})
\end{equation}
where $\epsilon \sim e^{-N/\xi}$ and $a$, $a^\dagger$ are the Dirac
fermion operators constructed from the boundary pair. Thus, the ground
state degeneracy is lifted by only an exponentially small splitting in
system size.

\section{General properties of quadratic fermionic Hamiltonians}
\label{sec:gener-prop-quadr}

We now step back and consider a generic quadratic fermionic
Hamiltonian:
\begin{equation}
 \label{eq:quad-ham}
 H(A) = \frac{i}{4}\sum_{j,k}A_{jk}c_jc_k
\end{equation}
where $A$ is a real, skew-symmetric matrix and the $c_j$ are Majorana
fermion operators. The normalization $\frac{i}{4}$ is convenient
because it has the property that 
\begin{equation}
 \label{eq:rep-so2n}
 [ -iH(A), -iH(B) ] = -i H\left( [A,B] \right)
\end{equation}
where $A,B \in \mathfrak{so}(2N)$, and $H(A), H(B)$ act on the Fock space
$\mathfrak{F}_N = \CC^{2^N}$. Thus $H(\cdot)$ provides a natural 
representation of $\mathfrak{so}(2N)$.

We now bring $H(A)$ to a canonical form:
\begin{equation}
 \label{eq:quad-canon}
 H_{\mathrm{canonical}} = \frac{i}{2} \sum_{k=1}^m \epsilon_k b_k'
 b_k'' = \sum_{k=1}^m \epsilon_k(\tilde{a}_k^\dagger \tilde{a}_k - \frac{1}{2})
\end{equation}
where $b_k'$, $b_k''$ are appropriate real linear combinations of the
original $c_j$ satisfying the same Majorana fermion commutation
relations and the $\tilde{a}_k, \tilde{a}_k^\dagger$ are the
annihilation and creation operators associated to the $b_k'$, $b_k''$
pair of Majoranas. This form for $H$ follows immediately from the
standard block diagonalization of real skew symmetric matrices
\begin{equation}
 \label{eq:10}
 A = Q \left( \begin{array}{ccccc}
0 & \epsilon_1 &&&\\
-\epsilon_1 & 0 &&&\\
&&0&\epsilon_2&\\
&&-\epsilon_2&0&\\
&&&&\cdots \end{array} \right) Q^T,~~ Q \in O(2N), \epsilon_k \ge 0
\end{equation}
From this form it is easy to check that the eigenvalues of $A$ are
$\pm i \epsilon_k$ and that the eigenvectors are the coefficients of
$c$ in $\tilde{a}_k$, $\tilde{a}_k^\dagger$. 

If some of the $\epsilon_k$ vanish, then we refer to the associated
fermions as \emph{zero modes}. In particular, these will lead to
ground state degeneracies, since occupation or nonoccupation of such
modes does not affect the energy. For the Majorana chain
of Eq. (\ref{eq:maj-ham}), we have
\begin{equation}
 \label{eq:7}
 A = \left( \begin{array}{cccccc}
0&v&&&&\\
-v&0&w&&&\\
&-w&0&v&&\\
&&-v&0&w&\\
&&&-w&0&\\
&&&&&\cdots \end{array} \right)
\end{equation}
We can find a vector $u$ such that $uA = 0$ by inspection:
\begin{equation}
 \label{eq:11}
 u = (1,0,\frac{v}{w},0,\left(\frac{v}{w}\right)^2,0,\cdots)
\end{equation}
This vector leads to a left boundary mode
\begin{equation}
 \label{eq:12}
 b_l = \sum u_k c_k
\end{equation}
while an analogous calculation starting at the right end will find a
right boundary mode $b_r$. These modes form a Majorana canonical pair,
leading to a two-fold degeneracy of the ground state of the
chain. Clearly, $u_k \sim e^{-k/\xi}$ falls off exponentially from the
edges of the chain with correlation length $\xi^{-1} =
\ln\left|\frac{w}{v}\right|$, as expected in section
\ref{sec:majorana-operators}.

\section{Why are the boundary modes robust?}
\label{sec:why-are-boundary}

In the simple case of a quadratic fermion Hamiltonian, we know that
the modes correspond to eigenvalues of a skew-symmetric real
matrix. These come in pairs $\pm i \epsilon$, in general, and the case
$\epsilon = 0$ is special. In particular, if the pair of Majoranas corresponding to a zero mode are physically well separated, we expect perturbations to have trouble lifting the boundary degeneracy. 

More generally, for interacting fermions, we can extend the symmetry
group $\ZZ_2$, generated by $P=P_{\mathrm{maj}}$, to a
\emph{non-commuting} algebra acting on the the ground state space
$\mf{L}$. First, in the noninteracting limit, at $v=0$, we define
\begin{equation}
 \label{eq:82}
 X = Y \prod_{k=1}^{j}(-i c_{2k-1} c_{2k})
\end{equation}
where $Y=c_{2j+1}$ is a local Majorana operator at site $2j+1$. A straightforward calculation shows that 
\begin{equation}
 \label{eq:85}
 XP = - PX
\end{equation}
and that $[H,X]=0$ so that the algebra generated by $X,P$ acts on
$\mf{L}$ nontrivially. We now allow $Y$ to vary as we adiabatically
turn on interactions and, so long as an energy gap is maintained, we
expect $Y$ to remain a local operator near $2j$, which we can separate
from the boundary by suitably large choice of $j$. That is, to find
$Y$, one needs to know the ground state or at least the structure of
the ground state near $2j$. This is a nontrivial operation but see \shortciteN{hastings:045141} for more details.

\chapter{The two-dimensional toric code}
\label{sec:two-dimens-mathbbz_2}

\renewcommand{\star}{\mathrm{star}}
\newcommand{\bdy}{\partial}

The toric code is an exactly solvable spin 1/2 model on the square
lattice. It exhibits a ground state degeneracy of $4^g$ when embedded
on a surface of genus $g$ and a quasiparticle spectrum with both
bosonic and fermionic sectors. Although we will not introduce it as
such, the model can be viewed as an Ising gauge theory at a
particularly simple point in parameter space (see
Sec.~\ref{sec:emergent-symmetry}). Many of the topological features of
the toric code model were essentially understood by  \shortciteN{PhysRevB.40.7133}, but they did not propose an
exactly solved model. A more detailed exposition of the toric code may
be found in \shortciteN{Kitaev20032}.

\begin{figure}
 \centering
 \vspace{0.5cm}
 \includegraphics[width=6cm]{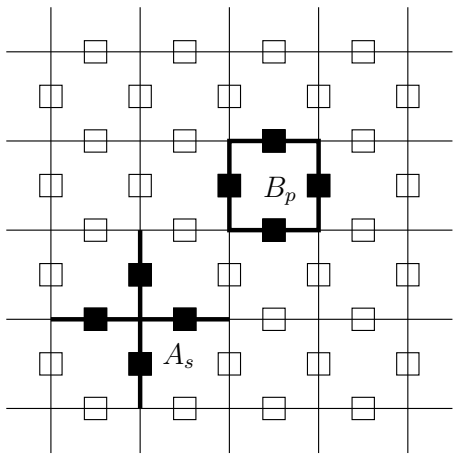}
 \vspace{0.5cm}

 \caption{A piece of the toric code. The spins live on the edges of the square lattice. The spins adjacent to a star operator $A_s$ and a plaquette operator $B_p$ are shown.}
 \label{fig:toric-code}
\end{figure}

We consider a square lattice, possibly embedded into a nontrivial
surface such as a torus, and place spins on the edges, as in
Fig.~\ref{fig:toric-code}. The Hamiltonian is given by
\begin{equation}
 \label{eq:toric-ham}
 H_T = -J_e \sum_s A_s - J_m \sum_p B_p
\end{equation}
where $s$ runs over the vertices (stars) of the lattice and $p$ runs
over the plaquettes. The star operator acts on the four spins
surrounding a vertex $s$,
\begin{equation}
 \label{eq:toric-star}
 A_s = \prod_{j\in \star(s)}\sigma_j^x
\end{equation}
while the plaquette operator acts on the four spins surrounding a
plaquette,
\begin{equation}
 \label{eq:toric-plaq}
 B_p = \prod_{j\in \bdy p} \sigma_j^z.
\end{equation}
Clearly, the $A_s$ all commute with one another, as do the
$B_p$. Slightly less trivially,
\begin{equation}
 \label{eq:14}
 A_s B_p = B_p A_s
\end{equation}
because any given star and plaquette share an even number of edges
(either none or two) and therefore the minus signs arising from the
commutation of $\sigma^x$ and $\sigma^z$ on those edges cancel. Since
all of the terms of $H_T$ commute, we expect to be able to solve it
term by term.

In particular, we will solve $H_T$ working in the $\sigma^z$
basis. Define classical variables $s_j = \pm 1$ to label the
$\sigma^z$ basis states. For each classical spin configuration
$\{s\}$, we can define the plaquette flux
\begin{equation}
 \label{eq:toric-vortex}
 w_p(s) = \prod_{j\in\bdy p} s_j.
\end{equation}
If $w_p = -1$, we say that there is a \emph{vortex} on plaquette $p$. 

\section{Ground states}
\label{sec:ground-states}

To find the ground states $\ket{\Psi}$ of $H_T$, we need to minimize
the energy, which means maximize the energy of each of the $A_s$ and
$B_p$ terms. The plaquette terms provide the condition
\begin{equation}
 \label{eq:15}
 B_p\ket{\Psi} = \ket{\Psi}
\end{equation}
which holds if and only if
\begin{equation}
 \label{eq:13}
 \ket{\Psi} = \sum_{\{s: w_p(s) = 1 ~\forall p\}} c_s \ket{s}
\end{equation}.
That is, the ground state contains no vortices. The group of star operators
act on the configurations $s$ by flipping spins. Thus, the star
conditions
\begin{equation}
 \label{eq:16}
 A_s\ket{\Psi} = \ket{\Psi}
\end{equation}
hold if and only if \emph{all of the $c_s$ are equal for each orbit
of the action of star operators}. In particular, if the spin flips of
$A_s$ are ergodic, as they are on the plane, all $c_s$ must be equal
and the ground state is uniquely determined.

\begin{figure}
 \centering
 \includegraphics[width=5cm]{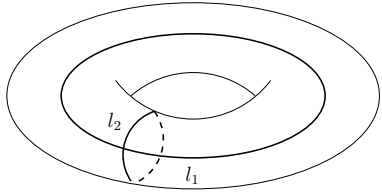}

 \caption{Large cycles on the torus.}
 \label{fig:torus-loops}
\end{figure}

On the torus, the star operators preserve the \emph{cohomology class}
of a vortex-free spin configuration. In more physical terms, we can
define conserved numbers given by the Wilson loop like functions
\begin{equation}
 \label{eq:18}
 w_l(s) = \prod_{j\in l} s_j,~~ l=l_1,l_2
\end{equation}
where $l_1$ and $l_2$ are two independent non-trivial cycles on the
square lattice wrapping the torus (Fig. \ref{fig:torus-loops}). Any
given star will overlap with a loop $l$ in either zero
or two edges and therefore $A_s$ preserves $w_l$. Since there are two
independent loops on the torus, each of which can have $w_l = \pm 1$,
there is a four-fold degenerate ground state:
\begin{equation}
 \label{eq:19}
 \ket{\Psi} = \sum_{\{s: w_p(s) = 1 ~\forall p\}} c_{w_{l_1}w_{l_2}} \ket{s}.
\end{equation}

\section{Excitations}
\label{sec:excitations}

\begin{figure}
 \centering
 \includegraphics[height=2.5cm]{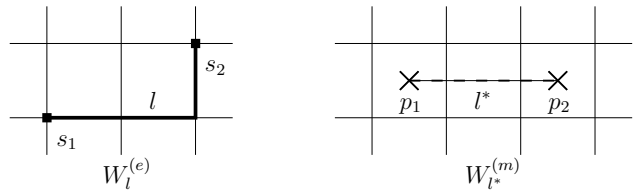}
 \caption{Electric and magnetic path operators.}
 \label{fig:toric-paths}
\end{figure}

The excitations of the toric code come in two varieties: the
\emph{electric charges} and \emph{magnetic vortices} of a $\ZZ_2$ gauge
theory. We will see this connection more explicitly later. In the
following, we restrict attention to the planar system for simplicity.

To find the electric charges, let us define the electric path operator
\begin{equation}
 \label{eq:toric-path-op}
 W_l^{(e)} = \prod_{j\in l} \sigma_j^z
\end{equation}
where $l$ is a path in the lattice going from $s_1$ to $s_2$ (see
Fig.~\ref{fig:toric-paths}). This operator clearly commutes with the
plaquette operators $B_p$ and with all of the star operators $A_s$
except for at the end points $s_1$ and $s_2$, where only one edge
overlaps between the star and the path and we have
\begin{equation}
 \label{eq:21}
 W_l^{(e)} A_{s_1} = - A_{s_1} W_l^{(e)}.
\end{equation}
Therefore, the state
\begin{equation}
 \label{eq:17}
 \ket{\Psi_{s_1,s_2}} = W_l^{(e)} \ket{\Psi_0},
\end{equation}
where $\ket{\Psi_0}$ is the planar ground state, is an eigenstate of
the Hamiltonian with excitations (charges) at $s_1$ and $s_2$ that
each cost energy $2 J_e$ to create relative to the ground state.

An analogous construction will find the magnetic vortices: we can
define a dual path operator
\begin{equation}
 \label{eq:20}
 W_{l^*}^{(m)} = \prod_{j\in l^*} \sigma_j^x
\end{equation}
where the path $l^*$ lies in the dual lattice (see
Fig. \ref{fig:toric-paths}) and goes from $p_1$ to $p_2$. In this
case, the stars $A_s$ all commute with $W_{l^*}^{(m)}$, as do all of
the plaquette operators $B_p$ except the two at the end points of
$l^*$, which anticommute. Thus, the $W_{l^*}^{(m)}$ operator creates a
pair of magnetic vortices on the plaquettes $p_1$ and $p_2$ at an
energy of $2 J_m$ each.


\chapter{Abelian anyons and quasiparticle statistics}
\label{sec:abelian-anyons}

Let us discuss what can possibly happen if we exchange two particles
in two dimensions. To ensure that particle statistics is well-defined,
we assume that there is no long-range interaction and that the phase
is gapped. If we drag two particles around one another adiabatically, 
\begin{displaymath}
 \label{eq:77}
  \includegraphics{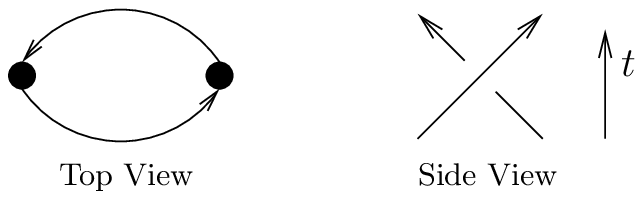}
\end{displaymath}
then we expect both dynamical
phase accumulation and a statistical effect due to the exchange. We
are well acquainted with this effect for everyday bosons and fermions,
for which:
\begin{equation}
 \label{eq:22}
 \begin{array}{lccl}
\textrm{Bosons:} & \ket{\Psi} & \mapsto & \ket{\Psi} \\
\textrm{Fermions:} & \ket{\Psi} & \mapsto & -\ket{\Psi} %
\end{array}
\end{equation}
where we have dropped the dynamical phase so as to focus on the
statistics. In both of these standard cases, a full rotation (two
exchanges), 
\begin{displaymath}
 \label{eq:78}
 \includegraphics{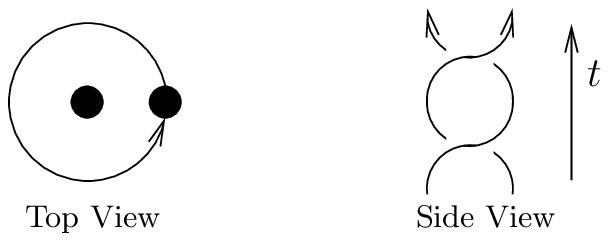}
\end{displaymath}
leaves $\ket{\Psi}$ unchanged.

\setlength{\unitlength}{1cm}
In principle, 
\begin{equation}
 \label{eq:23}
   \includegraphics{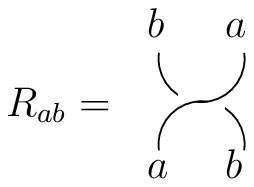}
\end{equation}
is an arbitrary phase factor or even an operator (\emph{braiding
 operator}). If the two particles are distinguishable ($a \ne b$),
then $R_{ab}$ does not have an invariant meaning, but the \emph{mutual
statistics} 
\begin{equation}
 \label{eq:24}
 \includegraphics{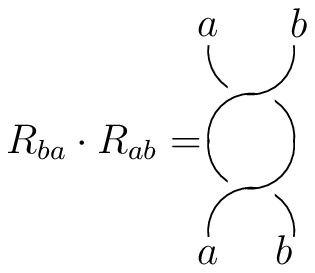}
\end{equation}
does.

Let us illustrate this in the toric code model. In section
\ref{sec:excitations} we found two kinds of quasiparticle excitations in the toric code: electric charges ($e$) and magnetic vortices ($m$). Since path
operators of the same type commute with one another, it is easy to
show that each of these are bosons. However, they have nontrivial
mutual statistics.

To calculate the mutual statistics, consider taking a charge $e$
around a vortex $m$.
\begin{displaymath}
\label{eq:79}
\includegraphics[scale=.8]{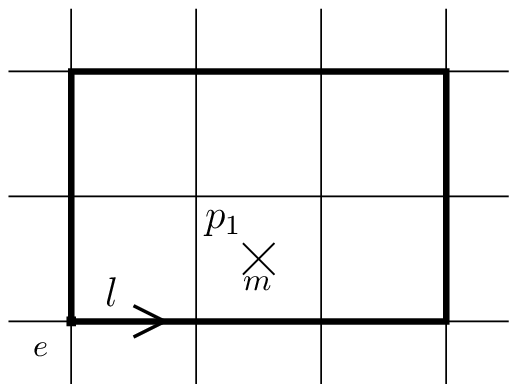}
\end{displaymath}
Let $\ket{\xi}$ be some state containing a magnetic vortex at
$p_1$. Under the full braiding operation,
\begin{eqnarray}
 \label{eq:25}
 \ket{\xi} & \mapsto & \left(\prod_{j\in l} \sigma_j^z\right)
 \ket{\xi} \nonumber\\
 & = & \left(\prod_{p \mathrm{\ inside\ } l} B_p\right) \ket{\xi}
\end{eqnarray}
where the second line is a Stokes' theorem like result relating the
product around a loop to the products of internal loops. Since 
\begin{equation}
 \label{eq:26}
 B_{p_1} \ket{\xi} = - \ket{\xi}
\end{equation}
for the plaquette $p_1$ containing the vortex, we have that
\begin{equation}
 \label{eq:27}
 \ket{\xi} \mapsto - \ket{\xi},
\end{equation}
or 
\begin{equation}
 \label{eq:28}
 \includegraphics{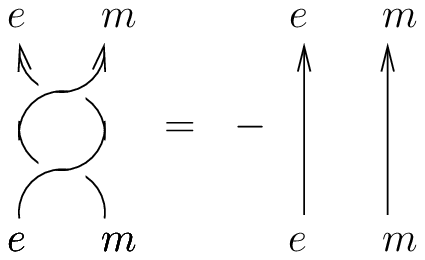}
\end{equation}

Using the bosonic self-statistics equations,
\begin{equation}
 \label{eq:29}
 \includegraphics{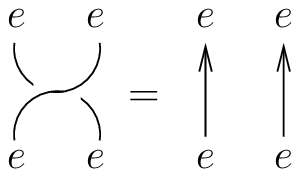}~~~~~~~~~
 \includegraphics{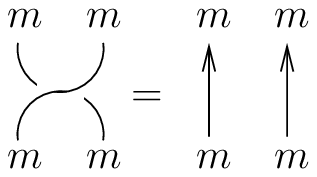}
\end{equation}
we can derive the nontrivial corollary that composite $e-m$ particles
are fermions:
\begin{equation}
 \label{eq:30}
 \includegraphics{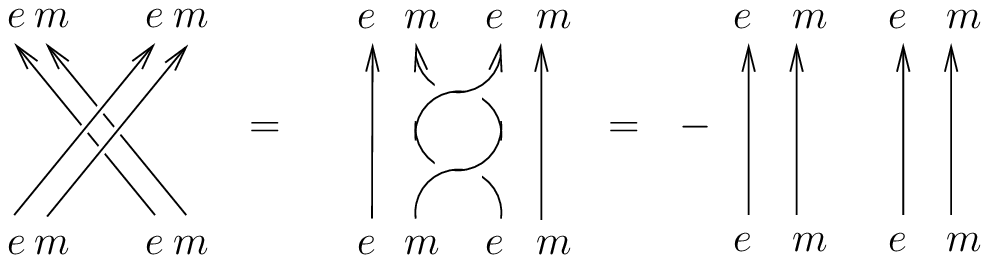}
\end{equation}

\section{Superselection sectors and fusion rules}
\label{sec:supers-sect-fusi}

Initially, we exhibited two kinds of bosonic excitations in the toric
code model (charges $e$ and vortices $m$) in the solution of the
Hamiltonian. After a bit of work, we discovered that a composite $e-m$
object has a meaningful characterization within the model as well, at
least in that it has fermionic statistics. This begs the question, how
many particle types exist in the toric code model and how can we
identify them?

We take an algebraic definition of a particle type: each type
corresponds to a \emph{superselection sector}, which is a
representation of the local operator algebra. In particular, we say
that two particles (or composite objects) are \emph{of the same
  type}
\begin{equation}
 \label{eq:31}
 a \sim b
\end{equation}
if $a$ can be transformed to $b$ by some operator acting in a finite
region. For example, in the toric code, two $e$-particles are
equivalent to having no particles at all,
\begin{equation}
 \label{eq:32}
 \includegraphics{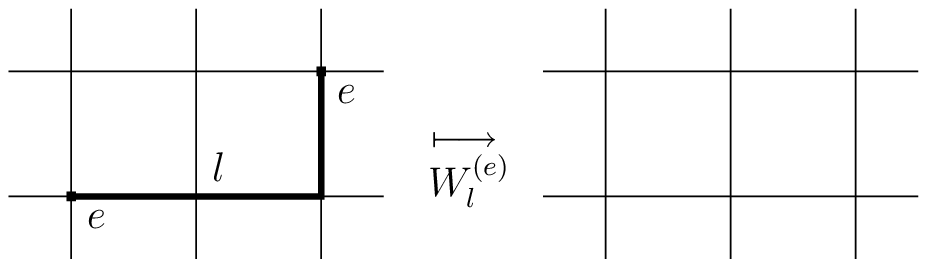}
\end{equation}
by acting with an appropriate, geometrically bounded electric path
operator $W^{(e)}_l$.

We introduce the notation
\begin{equation}
 \label{eq:33}
 e \times e = 1
\end{equation}
to represent the \emph{fusion rule} that two $e$-particles are
equivalent to the vacuum sector $1$. In the toric code, there are 4
superselection sectors: 
\begin{equation}
 \label{eq:34}
 1,~ e,~ m,~ \textrm{and}~ \epsilon=e\times m
\end{equation}
with the fusion rules:
\begin{equation}
 \label{eq:35}
 \begin{array}{ll}
e\times e = 1 & e\times m = \epsilon \\
m\times m = 1 & e \times \epsilon = m \\
\epsilon \times \epsilon = 1 & m \times \epsilon = e%
\end{array}
\end{equation}

\section{Mutual statistics implies degeneracy on the torus}
\label{sec:stat-impl-degen}

This is an argument due to \shortciteN{PhysRevLett.64.1995}. Suppose
that there are at least two particle types, $e$ and $m$ with $-1$
mutual statistics. Let us define an operator $Z$ acting on the ground
state in an abstract fashion (not refering to the actual model) which
creates an $e$ pair, wraps one particle around the torus and
annihilates the pair. In the toric code, this will be the path
operator $W_l^{(e)}=\prod_{j\in l}\sigma_j^z$ for a loop $l$ winding
one of the nontrivial cycles on the torus, but we need not know that
specifically.

We can define another operator $X$ that creates a pair of the other
type $m$ and winds around the other nontrivial cycle on the torus. But
now a bit of geometric introspection reveals that the combination,
\begin{equation}
 \label{eq:36}
 \includegraphics{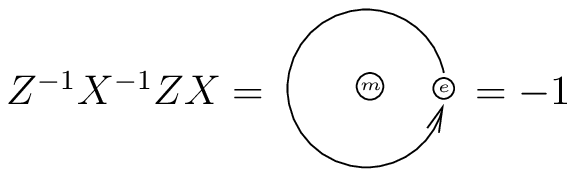}
\end{equation}
Thus, there are two non-commuting operators acting on the ground state
space $\mathfrak{L}$, and we conclude $\dim~\mathfrak{L} > 1$. In
fact, there are four such operators, each of the two particle types
can be moved around each of the two nontrivial cycles. Working out the
commutation relations of these operators implies that
$\dim~\mathfrak{L} = 4$.

\section{The toric code in a field: perturbation analysis}
\label{sec:pert-analys}

We now apply a magnetic field to the toric code that will
realistically allow the quasiparticles to hop and, unfortunately,
destroy its exact solvability (see \shortciteN{Tupitsyn:2008p5963}). To wit:
\begin{equation}
 \label{eq:37}
 H = - J_e \sum_s A_s - J_m \sum_p B_p - \sum_j\left(h_x \sigma_j^x +
   h_z \sigma_j^z\right)
\end{equation}
For example, with $h_x=0$ but $h_z \ne 0$, we can view the
perturbation as an electric path operator of length 1 on each
edge. Hence, it can cause charge pair creation and annihilation (at an
energy cost $\sim 4 J_e$) or hop existing charges by one lattice
displacement, at no cost. For small $h_z$ this provides a nontrivial
tight-binding dispersion to the charges,
\begin{equation}
 \label{eq:38}
 \epsilon(q) \approx 2 J_e - 2 h_z (\cos q_x + \cos q_y)
\end{equation}
but does not close the gap or lead to a change in the topological
degeneracy of the ground state in the thermodynamic limit. 

At large $h_z\gg J_e, J_m$, the model should simply align with the
applied field as a paramagnet. Clearly, in this limit the topological
degeneracy has been destroyed and we have a unique spin-polarized
ground state. The phase transition can be understood from the
topological side as a bose condensation of the charges, which
proliferate as $h_z$ increases.

The same argument is applicable if $h_x\gg J_e, J_m$. If $h_x$
increases while $h_z=0$, then vortices condense. However, the
high-field phase is just a paramagnet, so one can continuously rotate
the field between the $x$- and $z$-direction without inducing a phase
transition. Thus, the charge and vortex condensates are actually the
same phase! This property was first discovered by
\shortciteN{PhysRevD.19.3682} for a 3D classical $\ZZ_2$ gauge Higgs
model, where it appears rather mysterious.

\begin{figure}
  \vspace{-2.6cm}
  \begin{center}
    \includegraphics[width=8cm]{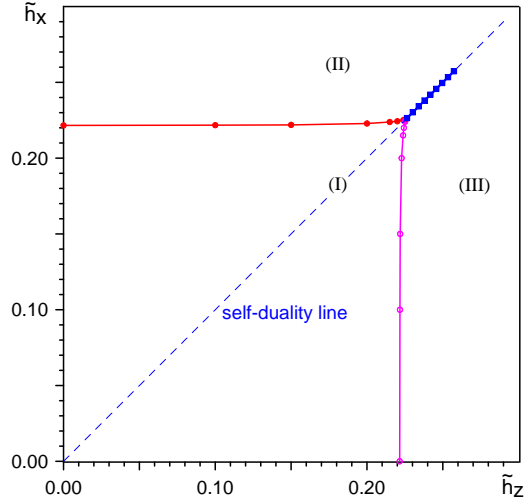}
  \end{center}
  \vspace{-2.6cm}
  \caption{%
    Numerically determined phase diagram of the toric code in a field
    from \protect\shortciteN{Tupitsyn:2008p5963}. (I) labels the
    topological phase, (II) and (III) the vortex and charge
    condensates (\emph{i.e.}  the paramagnetic phase). The numerics
    were done using discrete imaginary time with a rather large
    quantization step.%
  }
 \label{fig:toric-pd}
\end{figure}

\section{Robustness of the topological degeneracy}
\label{sec:robustn-topol-degen}

The splitting of the ground state levels due to virtual quasiparticle
tunneling is given by
\begin{equation}
 \label{eq:39}
 \delta E \sim \Delta e^{-L/\xi}
\end{equation}
This follows from the effective Hamiltonian
\begin{equation}
 \label{eq:40}
 H_{\mathrm{eff}} = -( t_{1Z} Z_1 + t_{2Z} Z_2 + t_{1X} X_1 + t_{2X}
 X_2 )
\end{equation}
where the $Z_i$, $X_i$ operators are the winding loop operators of
Sec.~\ref{sec:stat-impl-degen}. Physically, this is simply a
statement of the fact that the only way to act upon the ground state
is to wind quasiparticles around the torus. This is a process
exponentially suppressed in system size.

\section{Emergent symmetry: gauge formulation}
\label{sec:emergent-symmetry}

\newcommand{\tmu}{\tilde{\mu}}
\newcommand{\tsigma}{\tilde{\sigma}}
\newcommand{\tA}{\tilde{A}}
\newcommand{\tB}{\tilde{B}}

There are two ways to introduce symmetry operators in the perturbed
toric code model.

\begin{enumerate}
\item One can define \emph{loop operators} (e.g. $Z_1, Z_2, X_1,
  X_2$), the definition of which depends on the actual ground state of
  the perturbed Hamiltonian. This is similar to the definition of the
  operator Y in the 1D case of Sec.~\ref{sec:why-are-boundary}, which
  also requires detailed knowledge of the ground state.

\item One can exploit \emph{gauge invariance} by rewriting the model
  in a gauge invariant form. This can be done for any spin model by
  introducing redundancy. In this case, the symmetry does not depend
  on the model but is only manifest in the topological phase.
\end{enumerate}

We will take the second approach in order to avoid the difficulty of
defining the appropriate loop operators and also to introduce the
important gauge formulation of the model. To gauge the model we
proceed in steps:

\begin{enumerate}
\item Introduce one extra spin $\mu_v$ per vertex that always remains
 in the state
 \begin{equation}
   \label{eq:81}
   \frac{1}{\sqrt{2}}\left(\ket{\up}+\ket{\down}\right).
 \end{equation}
 This state is characterized by the condition
 \begin{equation}
   \label{eq:41}
   \mu_v^x \ket{\Psi} = \ket{\Psi},
 \end{equation}
 where $\mu_v^x$ is the Pauli spin matrix for the spin $\mu$ at vertex
 $v$.

\item Change spin operators from $\sigma_{uv}, \mu_v$ to
 $\tsigma_{uv},\tmu_v$ where we represent the classical value of each
 old spin $s_{uv}$ as $\tilde{m}_u \tilde{s}_{uv} \tilde{m}_v$. Here
 $s_{uv}$ is the spin on the edge connecting $u$ and $v$ and $m_u$,
 $m_v$ are the classical values of the new spins (\emph{i.e.} the
 labels in the $\mu^z$ basis).
\end{enumerate}

Thus the complete transformation is given by
\begin{eqnarray}
 \label{eq:42}
 \sigma_{uv}^z &=& \tmu^z_u \tsigma^z_{uv} \tmu^z_v \nonumber\\
 \sigma_{uv}^x &=& \tsigma_{uv}^x \nonumber\\
 \mu_u^z &=& \tmu_u^z \nonumber\\
 \mu_u^x &=& \tmu_u^x \prod_{j\in \star(u)}\tsigma_j^x = \tmu_u^x \tA_u
\end{eqnarray}
and the constraint Eq. (\ref{eq:41}) becomes the standard $\ZZ_2$
gauge constraint:
\begin{eqnarray}
 \label{eq:z2-gauge-constraint}
 \tmu_u^x \tA_u \ket{\Psi} = \ket{\Psi}.
\end{eqnarray}

On states satisfying the gauge constraint
Eq. (\ref{eq:z2-gauge-constraint}), $A_u = \tA_u =
\tmu_u^x$. Therefore, 
\begin{eqnarray}
 \label{eq:43}
 A_u\ket{\Psi} = \tmu_u^x \ket{\Psi}
\end{eqnarray}
and we can rewrite the Hamiltonian as
\begin{eqnarray}
 \label{eq:44}
 H = -J_e\sum_v \tmu_v^x - J_m\sum_p \tB_p -
 \sum_{\left<u,v\right>}\left(h_x \tsigma_{uv}^x + h_z \tmu_u
   \tsigma_{uv}^z \tmu_v\right)
\end{eqnarray}
subject to the gauge constraint.

Viewed as a standard $\ZZ_2$ gauge theory, the protected topological
degeneracy of the ground state is physically familiar as the protected
degeneracy associated with the choice of flux threading the
$2g$ holes of the genus $g$ surface.

\chapter{The honeycomb lattice model}
\label{sec:honeyc-latt-model}

\begin{figure}
 \centering
 \includegraphics{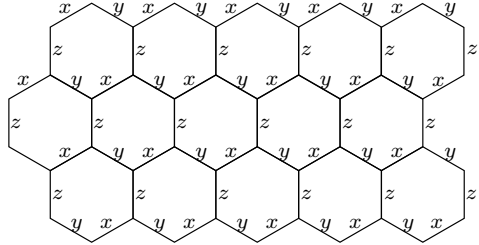}
 \caption{The honeycomb model has spins living on the vertices of a
   honeycomb lattice with nearest neighbor interactions that are
   link-orientation dependent. $x$-links have $\sigma^x\sigma^x$
   interactions, $y$-links have $\sigma^y\sigma^y$ interactions and
   $z$-links have $\sigma^z\sigma^z$ interactions.}
 \label{fig:honeycomb-lattice}
\end{figure}

We now investigate the properties of another exactly solvable spin
model in two dimensions, the \emph{honeycomb lattice model}. This
model exhibits a number of gapped phases that are perturbatively
related to the toric code of the previous sections. Moreover, in the
presence of time-reversal symmetry breaking terms, a new topological
phase arises with different topological properties, including
nontrivial \emph{spectral Chern number}. An extended treatment of the
properties of this model with much greater detail can be found in
\shortciteN{Kitaev20062}.

In the honeycomb lattice model, the degrees of freedom are spins
living on the vertices of a honeycomb lattice with nearest neighbor
interactions. The unusual feature of this model is that the
interactions are link orientation dependent (see
Fig. \ref{fig:honeycomb-lattice}).  The Hamiltonian is
\begin{equation}
 \label{eq:45}
 H = -J_x \sum_{x~\mathrm{ links}}\sigma_j^x \sigma_k^x 
 -J_y \sum_{y~\mathrm{ links}}\sigma_j^y \sigma_k^y 
 -J_z \sum_{z~\mathrm{ links}}\sigma_j^z \sigma_k^z 
\end{equation}
We might expect this model to be integrable because $[H,W_p]=0$ for an
extensive collection of plaquette operators
\begin{equation}
 \label{eq:46}
\includegraphics{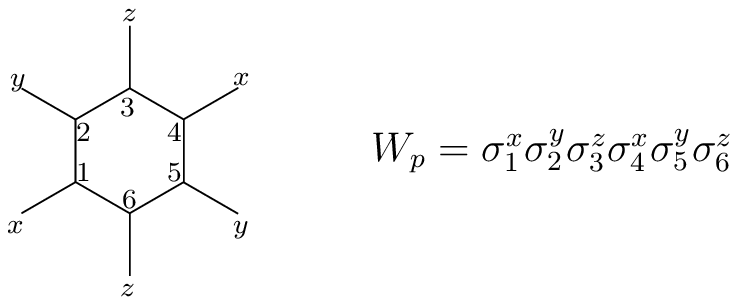}
\end{equation}
where the spins and labels follow from the figure for each
plaquette. Unfortunately, this is not quite enough: there are two
spins but only one constraint per hexagon so that half of each spin
remains unconstrained. In fact, the remaining degrees of freedom are
Majorana operators!

\section{A (redundant) representation of a spin by 4 Majorana operators}
\label{sec:redund-repr-spin}

\newcommand{\ts}{\tsigma}

We consider a collection of four Majorana operators $c, b^x, b^y$ and
$b^z$ that act on the 4-dimensional Fock space
$\mathfrak{F}$. We define the following three operators 
\begin{eqnarray}
 \label{eq:48}
 \ts^x &=& i b^x c \nonumber\\
 \ts^y &=& i b^y c \nonumber\\
 \ts^y &=& i b^y c.
\end{eqnarray}
These operators do not obey the spin algebra relations on the full
Fock space, but we clearly have two extra dimensions of wiggle
room. In fact, the physical state space is identified with a
two-dimensional subspace $\mf{L}\subset \mf{F}$ given by
the constraint
\begin{eqnarray}
 \label{eq:49}
 D\ket{\Psi} = \ket{\Psi},~~~~\textrm{where }D=b^xb^yb^zc
\end{eqnarray}
Within $\mf{L}$, the $\ts^\alpha$ act as $\sigma^\alpha$ act on
the actual spin. Of course, $\ts^\alpha$ also act on
$\mf{L}^\perp$, but we can ignore these states by enforcing the 
constraint.

To be careful, we need to check two consistency conditions:
\begin{enumerate}
\item $\ts^\alpha$ preserves the subspace $\mf{L}$, which follows from
 $[\ts^\alpha, D] = 0$.
\item The $\ts^\alpha$ satisfy the correct algebraic relations
 when restricted to $\mf{L}$. For example,
 \begin{equation}
   \label{eq:50}
   \ts^x \ts^y \ts^z = (ib^xc)(ib^yc)(ib^zc) = i^3(-1)b^xb^yb^zc^3 =
   i D = i
 \end{equation}
 where the last equality only holds in the physical subspace $\mf{L}$.
\end{enumerate}

\section{Solving the Honeycomb Model using Majoranas}
\label{sec:solv-honeyc-model}

\begin{figure}
 \centering
 \includegraphics[scale=.8]{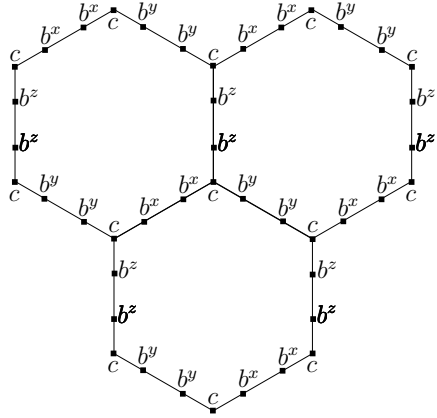}
 \caption{Majorana representation of honeycomb model.}
 \label{fig:honeycomb-maj}
\end{figure}

We now use the Majorana representation of spins just introduced to
rewrite each spin of the entire honeycomb model as in
Fig.~\ref{fig:honeycomb-maj}. This will greatly expand the
$2^N$-dimensional Hilbert space to the Fock space $\mf{F}$ of
dimension $2^{2N}$, but the physical space $\mf{L} \subset \mf{F}$ is
fixed by the gauge condition
\begin{equation}
 \label{eq:51}
 D_j\ket{\Psi} = \ket{\Psi} ~~~\textrm{for all } j
\end{equation}
where $D_j = b_j^x b_j^y b_j^z c$. We define a projector onto
$\mf{L}$ by
\begin{equation}
 \label{eq:52}
 \Pi_{\mf{L}} = \prod_j\left(\frac{1+D_j}{2}\right)
\end{equation}
In the Majorana representation, the Hamiltonian (\ref{eq:45}) becomes
\begin{eqnarray}
 \label{eq:53}
 \tilde{H} &=& \frac{i}{4}\sum_{\left<j,k\right>}\hat{A}_{jk}c_jc_k \nonumber\\
 \hat{A}_{jk} &=& 2 J_{\alpha(j,k)} \hat{u}_{jk} \nonumber\\
 \hat{u}_{jk} &=& i b_j^{\alpha(j,k)} b_k^{\alpha(j,k)}
\end{eqnarray}
where $\alpha(j,k) = x,y,z$ is the direction of the link between $j$
and $k$. 

We have suggestively written the Hamiltonian $\tilde{H}$ as if it were
a simple quadratic fermion Hamiltonian as in
Sec.~\ref{sec:gener-prop-quadr}, but of course $\hat{A}_{jk}$ is
secretly an operator rather than a real skew-symmetric
matrix. However, each operator $b_j^\alpha$ enters only one term of
the Hamiltonian and therefore $\hat{u}_{jk}$ commute with each other
and with $\tilde{H}$! Thus, we can fix $u_{jk}=\pm 1$, defining an
orthogonal decomposition of the full Fock space:
\begin{equation}
 \label{eq:54}
 \mf{F} = \bigoplus_{u}\mf{F}_u,~~\textrm{ where } \ket{\Psi} \in
 \mf{F}_u\textrm{ iff } \hat{u}_{jk}\ket{\Psi} =
 u_{jk}\ket{\Psi}~~\forall~j,k
\end{equation}

Within each subspace $\mf{F}_u$, we need to solve the quadratic
Hamiltonian
\begin{eqnarray}
 \label{eq:55}
 \tilde{H}_u &=& \frac{i}{4}\sum_{\left<j,k\right>}A_{jk}c_jc_k
 \nonumber\\
 A_{jk} &=& 2 J_{\alpha(j,k)} u_{jk}
\end{eqnarray}
which we know how to do in principle. On the other hand, the integrals
of motion $W_p$ (the hexagon operators) define a decomposition of the
physical subspace $\mf{L}$ labeled by the eigenvalues $w_p = \pm 1$:
\begin{equation}
 \label{eq:56}
 \mf{L} = \bigoplus_w \mf{L}_w,~~\textrm{ where } \ket{\Psi} \in
 \mf{L}_w\textrm{ iff } W_p\ket{\Psi} = w_p \ket{\Psi}~~\forall~p
\end{equation}
We can relate these two decompositions by expressing $W_p$ in the Majorana representation and noting that within the physical subspace
\begin{equation}
 \label{eq:59}
 \tilde{W}_p  = \prod_{\left<j,k\right>\in\bdy p} \hat{u}_{jk}
\end{equation}
Thus, we find
\begin{equation}
 \label{eq:57}
 \mf{L}_w = \Pi_{\mf{L}} \mf{F}_u
\end{equation}
where $w_p = \prod_{(j,k)\in\bdy p}u_{jk}$. 

So we have a procedure for finding the ground state of the honeycomb model:
\begin{enumerate}
\item Fix $w_p = \pm 1$ for all $p$.

\item Find $u_{jk}$ satisfying
 \begin{equation}
   \label{eq:60}
   w_p = \prod_{(j,k)\in\bdy p}u_{jk}.
 \end{equation}
 There is a small subtletly here in that $u_{jk} = - u_{kj}$ so we must be careful about ordering. We can consistently take $j$ in the even sublattice of the honeycomb and $k$ in the odd sublattice in equation (\ref{eq:60}).

\item Solve for the ground state of the quadratic Hamiltonian (\ref{eq:55}), finding the energy $E(w)$.

\item Project the found state onto the physical subspace (\emph{i.e.} symmetrize over gauge transformations).

\item Repeat for all $w$; pick the $w$ that minimizes the energy.
\end{enumerate}

\begin{figure}
 \centering
 \includegraphics{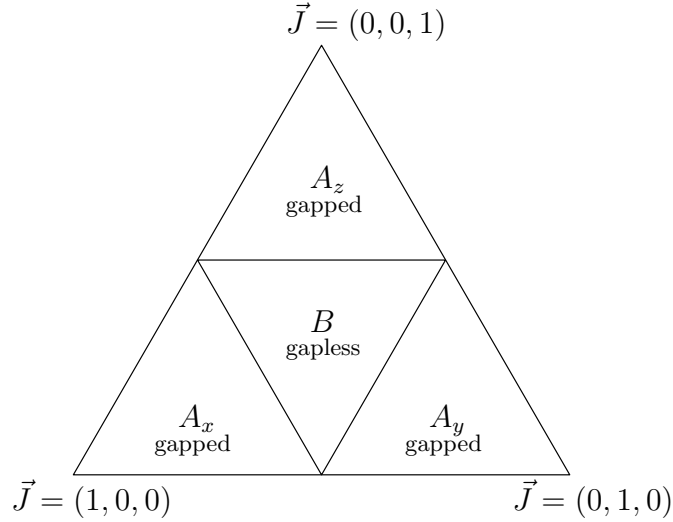}
 \caption{Phase diagram of honeycomb model. This is a slice through
   the positive octant in $\vec{J}$ coupling space along the
   $J_x+J_y+J_z=1$ plane. The other octants are analogous.}
 \label{fig:honeycomb-pd}
\end{figure}

If there were no further structure to $E(w)$, this would be an
intractable search problem in the space of $w_p$. Fortunately, due to
a theorem by Lieb~\citeyear{PhysRevLett.73.2158}, the ground state has
no vortices. That is,
\begin{equation}
 \label{eq:61}
 E(w) = \textrm{min    if } w_p = 1~\forall~p
\end{equation}
Using this choice of $w_p$, it is easy to solve the model and produce
the phase diagram of Fig.~\ref{fig:honeycomb-pd}. The gapless phase
has two Dirac points in the fermionic spectrum.

\section{Fermionic spectrum in the honeycomb lattice model}
\label{sec:ferm-spectr-honeyc}

\begin{figure}
 \centering
 \includegraphics{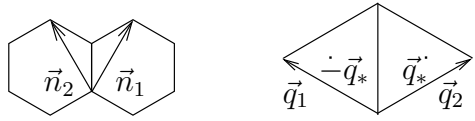}
 \caption{Direct and reciprocal lattices of the honeycomb. The points $\pm\vec{q}_*$ are the two Dirac points of the gapless phase B.}
 \label{fig:honeycomb-momspace}
\end{figure}

We just need to diagonalize the Hamiltonian
\begin{eqnarray}
 \tilde{H}_u &=& \frac{i}{4}\sum_{\left<j,k\right>}A_{jk}c_jc_k
 \nonumber\\
 A_{jk} &=& 2 J_{\alpha(j,k)} u_{jk} \nonumber\\
 u_{jk} &=&\left\{\begin{array}{cl}
+1&\textrm{ if } j\in\textrm{even sublattice}\\
-1&\textrm{ otherwise} \end{array}\right.
\end{eqnarray}
This is equivalent to finding the eigenvalues and eigenvectors of the matrix $iA$. Since the honeycomb lattice has two sites per unit cell, by applying the Fourier transform we get a $2\times 2$ matrix $A(\vec{q})$:
\begin{eqnarray}
 \label{eq:58}
 iA(\vec{q}) & = & \left( \begin{array}{cc}
0&i f(\vec{q}) \\
-i f(\vec{q}) & 0 \end{array} \right) \nonumber\\
 \epsilon(\vec{q}) &=& \pm|f(\vec{q})|
\end{eqnarray}
where $f(\vec{q})$ is some complex function that depends on the
couplings $J_x, J_y, J_z$. In the gapless phase (phase B in
Fig.~\ref{fig:honeycomb-pd}), $f(\vec{q})$ has two zeros which
correspond to Dirac points (see Fig.~\ref{fig:honeycomb-momspace}). At
the transition to phase A, the Dirac points merge and disappear.

\section{Quasiparticle statistics in the gapped phase}
\label{sec:quas-stat-gapp}

It appears that there are two particle types: fermions and vortices (hexagons
with $w_p = -1$). The vortices are associated with a $\ZZ_2$ gauge field,
where $u_{jk}$ plays the role of vector potential. Taking a fermion around a
vortex results in the multiplication of the state by $-1$ (compared to the
no-vortex case). However, the details such as the fusion rules are not
obvious.

\begin{figure}
 \centering
\includegraphics{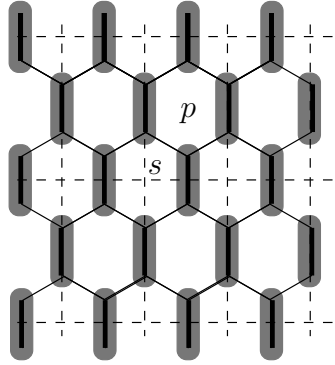}
\caption{The vertical dimers on the honeycomb lattice themselves form
 the edges of a (dashed) square lattice. The plaquettes of alternate
 rows of the hexagonal lattice correspond to the stars and plaquettes
 of the square lattice.  This is \emph{weak breaking of translational
   symmetry.}}
 \label{fig:honeycomb-dimer}
\end{figure}

Let us look at the model from a different perspective. If $J_x=J_y=0,
J_z>0$, the system is just a set of dimers (see
Fig.~\ref{fig:honeycomb-dimer}). Each dimer can be in two states:
$\up\up$ and $\down\down$. The other two states have $2 J_z$ higher
energy. Thus, the ground state is highly degenerate.

If $J_x, J_y \ll J_z$, we can use perturbation theory relative to the noninteracting dimer point. Let us characterize each dimer by an effective spin:
\begin{eqnarray}
 \label{eq:62}
 \ket{\Uparrow} = \ket{\up\up};~~~~~\ket{\Downarrow} = \ket{\down\down}.
\end{eqnarray}
At 4th order of perturbation theory, we get:
\begin{eqnarray}
 \label{eq:63}
 H^{(4)}_{\mathrm{eff}} = \mathrm{const} - \frac{J_x^2 J_y^2}{16 J_z^3}\sum_p Q_p
\end{eqnarray}
where $p$ runs over the square plaquettes of the dimer lattice (see Fig.~\ref{fig:honeycomb-dimer}) and  
\begin{eqnarray}
 \label{eq:64}
 Q_p = \sigma_{p_1}^y\sigma_{p_2}^x\sigma_{p_3}^y\sigma_{p_4}^x
\end{eqnarray}
is a plaquette operator on the effective spin space $\ket{\Uparrow}$,
$\ket{\Downarrow}$. By adjusting the unit cell and rotating the spins,
we can reduce this Hamiltonian to the toric code!

The vertices and plaquettes of the new lattice correspond to
alternating rows of hexagons. Thus, vortices on even rows belong to
one superselection sector and vortices on odd rows to the other. It is
impossible to move a vortex from an even row to an odd row by a local
operator without producing other particles (\emph{e.g.} fermions). The
fermions and $e-m$ pairs belong to the same superselection sector,
$\epsilon$, though these are different physical states.

\section{Nonabelian phase}
\label{sec:non-abelian-phase}

In the gapless phase B, vortex statistics are not
well-defined. However, a gap can be opened by applying a perturbation
that breaks the time-reversal symmetry, such as a magnetic
field. Unfortunately the honeycomb model in a field is not exactly
solvable. \shortciteN{yao:247203} studied an exactly solvable spin
model where the time-reversal symmetry is spontaneously broken, but we
will satisfy ourselves by introducing a T-breaking next nearest
neighbor interaction on the fermionic level (which can be represented
by a 3-spin interaction in the original spin language).

\begin{figure}
 \centering
 \includegraphics{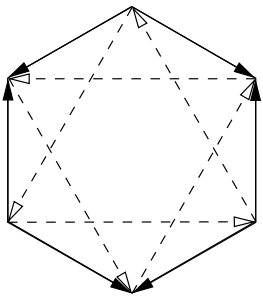}
 \caption{Picture of chiral interaction matrix $A_{jk}$. Forward
   arrows correspond to positive entries in the skew-symmetric real
   matrix $A_{jk}$. Solid arrows are the interactions of the original
   honeycomb model; dashed arrows give the time-reversal symmetry
   breaking perturbation.}
 \label{fig:honeycomb-trs}
\end{figure}

Written in terms of Majorana fermions, we consider the Hamiltonian
\begin{equation}
 \label{eq:47}
 H = \frac{i}{4} \sum_{\left<j,k\right>}A_{jk}c_jc_k
\end{equation}
where $A_{jk}$ now has chiral terms connecting Majoranas beyond nearest
neighbor in the honeycomb lattice (see
Fig.~\ref{fig:honeycomb-trs}). After Fourier transforming, we find
\begin{equation}
 \label{eq:65}
 i A(\vec{q}) = \left(\begin{array}{cc}
\Delta(\vec{q}) & i f(\vec{q}) \\
- i f(\vec{q}) & - \Delta(\vec{q}) \end{array} \right)
\end{equation}
with the massive dispersion relation
\begin{equation}
 \label{eq:66}
 \epsilon(\vec{q}) = \pm\sqrt{f(\vec{q})^2+\Delta(\vec{q})^2}.
\end{equation}

Within this massive phase, we will find nontrivial topological
invariants of the quasiparticle spectrum. Let $i A(\vec{q})$ be a
nondegenerate Hermitian matrix that continously depends on
$\vec{q}$. In our case, $A$ acts in $\CC^2$, but in general it can be
$\CC^n$ for any $n$. Let us keep track of the ``negative eigenspace''
of $i A(\vec{q})$: the subspace $\mf{L}(\vec{q}) \subseteq \CC^n$
spanned by eigenvectors corresponding to negative eigenvalues. For
matrix (\ref{eq:65}), $\dim~\mf{L}(\vec{q}) = 1$. This defines a map
$F$ from momentum space (the torus) to the set of $m$-dimensional
subspaces in $\CC^n$. More formally:
\begin{equation}
 \label{eq:67}
 F:~\mathbb{T}^2 \longrightarrow U(n)/U(m)\times U(n-m)
\end{equation}
This map $F$ may have nontrivial topology.

In the honeycomb model with T-breaking, we have $n=2$, $m=1$ and $U(2)
/ U(1)\times U(1) = \CC P^1 = S^2$ is the unit sphere. Thus, $F:
\mathbb{T}^2\longrightarrow S^2$ and for the matrix $i A(\vec{q})$ of
Eq.~(\ref{eq:65}), $F$ has degree 1. That is, the torus wraps around
the sphere once. More abstractly, $\mf{L}(\vec{q})$ defines a complex
vector bundle over the momentum space $\mathbb{T}^2$. This has an
invariant Chern number $\nu$, which in this case is $\nu = 1$.

What is the significance of the spectral Chern number? It is known to
characterize the integer quantum Hall effect, where it is known as the
``TKNN invariant''. For a Majorana system, there is no Hall effect
since particles are not conserved. Rather, the spectral Chern number
determines the number of chiral modes at the edge:
\begin{equation}
 \label{eq:68}
 \nu = (\textrm{\# of left-movers}) - (\textrm{\# of right-movers}).
\end{equation}

\section{Robustness of chiral modes}
\label{sec:robustn-chir-modes}

A chiral edge mode may be described by its Hamiltonian:
\begin{equation}
 \label{eq:69}
 H_{\mathrm{edge}} = \frac{i v}{4}\int \hat{\eta}(x) \partial_x \hat{\eta}(x) dx
\end{equation}
where $\hat{\eta}(x)$ is a real fermionic field. That is,
\begin{equation}
 \label{eq:70}
 \hat{\eta}(x) \hat{\eta}(y) + \hat{\eta}(y) \hat{\eta}(x) = 2 \delta(x - y).
\end{equation}
At temperature $T$, each mode carries energy current
\begin{equation}
 \label{eq:71}
 I_1 = \frac{\pi}{24}T^2.
\end{equation}

The easiest explanation of this is a straightforward 1-D fermi gas
calculation:
\begin{eqnarray}
 \label{eq:72}
 I_1 &=& v\int_0^\infty n(q) \epsilon(q) \frac{dq}{4\pi} \nonumber\\
 &=& \frac{1}{2\pi}\int_0^\infty\frac{\epsilon d\epsilon}{1+e^{\epsilon/T}} \nonumber\\
 &=& \frac{\pi}{24}T^2
\end{eqnarray}
However, it is useful to reexamine this current using conformal field
theory (CFT), in order to understand better why the chiral modes are
robust. We consider a disc of $B$ phase extended into imaginary time
at temperature $T$. That is, we have a solid cylinder with top and
bottom identified:
\begin{displaymath}
 \label{eq:86}
 \includegraphics{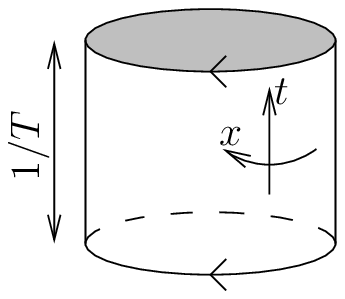}
\end{displaymath}
We have obtained a solid torus whose surface is a usual torus. The partition function is mostly determined by the surface.

Let the spatial dimensions be much greater than $\frac{1}{T}$. From
this point of view, the cylinder looks more like:
\begin{displaymath}
 \label{eq:87}
 \includegraphics{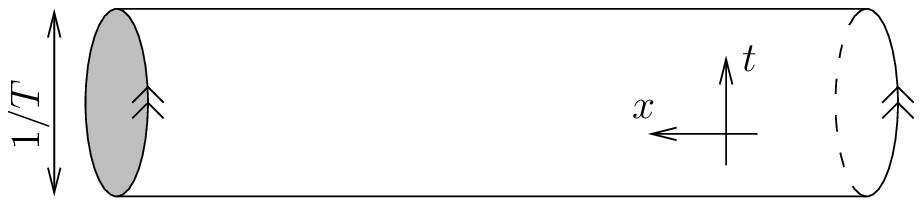}
\end{displaymath}
According to the usual CFT arguments, we have
\begin{eqnarray}
 \label{eq:73}
 Z &\sim& q^{\frac{c}{24}}\bar{q}^{\frac{\bar{c}}{24}} \nonumber\\
 q &=& e^{2 \pi i \tau} \nonumber\\
 \tau &=& i \frac{L T}{v} + \textrm{twist}
\end{eqnarray}
Twisting the torus changes the partition function by:
\begin{eqnarray}
 \label{eq:74}
 \tau &\mapsto&\tau+1\nonumber\\
 Z & \mapsto & Z e^{2\pi i \frac{c-\bar{c}}{24}}.
\end{eqnarray}
On the other hand, the twist parameter ($\Re \tau$) couples to the some
component of the energy-momentum tensor, namely, $T_{xt}$, which
corresponds to the energy flow. This relation implies that
\begin{eqnarray}
 \label{eq:75}
 I & = & \frac{\pi}{12}(c-\bar{c}) T^2.
\end{eqnarray}
The chiral central charge, $c-\bar{c}$, does not depend on the boundary conditions. Indeed, the energy current on the edge cannot change because the energy cannot go into the bulk.

\bibliographystyle{OUPnamed}
\bibliography{top-phases}

\end{document}